\documentstyle[preprint,prb,aps,epsf]{revtex}
\begin{document}
\draft
\title{ Effects of an in-plane magnetic field on $c$-axis sum rule and
superfluid density in high-$T_{c}$ cuprates}
\author{Wonkee Kim and J. P. Carbotte}
\address{Department of Physics and Astronomy, McMaster University,
Hamilton, Ontario, Canada L8S~4M1}
\maketitle
\begin{abstract}
In layered cuprates, the application of an in-plane
magnetic field $({\bf H})$ changes the $c$-axis optical sum rule and
superfluid density $\rho_{s}$. For pure incoherent $c$-axis coupling, 
${\bf H}$ has no effect on either quantities
but it does if an additional coherent component is present.
For the coherent contribution, 
different characteristic variations on ${\bf H}$ and
on temperature result from the constant part $(t_{\perp})$ of the 
hopping matrix element and from the part $(t_{\phi})$ which
has zero on the diagonal of the Brillouin zone. 
Only the 
constant part $(t_{\perp})$ leads to a dependence on the direction of
${\bf H}$ as well as on its magnitude.
\end{abstract}

A conventional $c$-axis optical sum rule\cite{sum} has been observed
in some high-$T_{c}$ cuprates such as optimally doped YBCO,\cite{basov}
while in others, such as underdoped YBCO, it is violated. The sum rule is 
formulated in terms of the missing area at finite energy $\omega$ under 
the real part of the optical conductivity when the sample
becomes superconducting. In the conventional case this missing spectral
weight $(\Delta N)$ shows up as the superfluid density $(\rho_{s})$
at zero energy and $\Delta N/\rho_{s}=1$. 
Underdoped YBCO exhibits a pseudo gap above the superconducting
critical temperature $(T_{c})$.
This is taken as an indication of non-Fermi liquid
in-plane behavior and $\Delta N/\rho_{s}\simeq1/2$. 
Even for an in-plane Fermi liquid, deviations\cite{kim}
from one result for $\Delta N/\rho_{s}$ if the $c$-axis coupling
is incoherent, {\it i.e.} does not conserve in-plane momentum. 
On the other hand, the conventional value of
$\Delta N/\rho_{s}$ observed in optimally doped YBCO is
consistent with an in-plane Fermi liquid and coherent $c$-axis
dynamics\cite{kim} whether or not the hopping amplitude depends on in-plane
momentum. It is clear, therefore, that such measurements can yield
important information on the dynamics of the charge carriers in the
cuprates.

Information on the nature of the interlayer coupling can also be obtained 
from other measurements. For example, 
it was recognized\cite{radtke} that the temperature $(T)$
variation of the $c$-axis penetration depth would mirror the observed
in-plane linear $T$ dependence, if the $c$-axis coupling is coherent
with a constant matrix element $t_{\perp}$.
However, studies\cite{xiang1,sanderman}
of atomic overlapps between CuO$_{2}$ planes through
which $c$-axis tunneling occurs, have shown that the chemistry involved
leads to a matrix element $t(\phi)$ which 
has $d$-wave symmetry and is given by 
$t_{\phi}\cos^{2}(2\phi)$, where $\phi$ 
gives the direction of the momentum
${\bf k}$ in the 2-dimensional CuO$_{2}$ Brillouin zone. In general,
there could be a contribution from
both a constan $t_{\perp}$ and $t(\phi)$. For the case of
$t(\phi)$, the low $T$ dependence of the resulting $\rho_{s}(T)$
for a $d$-wave superconductor is $T^{5}$ 
(See Ref.\cite{xiang1}). Such a dependence on $T$
has been observed in a recent 
experiment\cite{gaifullin} on Bi2212 
indicating that $t(\phi)$ is the dominant coherent matrix element.
While this experiment 
favors a $T^{5}$, some small contribution from a constant $t_{\perp}$
and from some incoherent transfer, 
which gives a $T^{3}$ law,\cite{radtke,hirschfeld}
cannot be ruled out. In fact, there is 
strong evidence from the $c$-axis quasiparticle tunneling work\cite{latyshev} 
on mesa junctions
of Bi2212 that there is a significant $t_{\perp}$
part in the coherent matrix element.\cite{kim2}

In this paper we investigate the effect of an in-plane magnetic field
${\bf H}$ on the $c$-axis sum rule and on the corresponding 
superfluid density. The aim is to see if additional information on 
the nature of the
$c$-axis coupling can be obtained from such experiments. We consider a general
coherent $c$-axis coupling matrix element 
$\left[t_{\perp}({\bf k})=t_{\perp}+t_{\phi}\cos^{2}(2\phi)\right]$
as well as impurity mediated transfer from plane to plane.

We begin with the Hamiltonian ${\cal H}=H_{0}+H_{c}$, where $H_{0}$ describes
a $d$-wave superconductor in the plane and
$H_{c}$ gives the $c$-axis coupling from plane to plane
$H_{c}=\sum_{\sigma,{\bf k},{\bf p}}t_{{\bf k}-{\bf p}}
\left[C^{+}_{1\sigma}({\bf k})
C_{2\sigma}({\bf p})+h.c.\right]$,
where $C^{+}_{i\sigma}({\bf k})$ creates an electron in plane $i$ of momentum
{\bf k} and spin $\sigma$, and $t_{{\bf k}-{\bf p}}$ is the hopping 
amplitude between plane $1$ and $2$. For coherent
coupling, $t_{{\bf k}-{\bf p}}=\delta_{{\bf k}-{\bf p}}t_{\perp}({\bf k})$.
For impurity assisted hopping
$t_{{\bf k}-{\bf p}}=V_{{\bf k}-{\bf p}}$, where $V_{{\bf k}-{\bf p}}$
is the impurity potential, and in-plane momentum is not conserved once
a configuration average over impurities is carried out.

Consideration of $c$-axis electrodynamics leads directly to
a relation for the superfluid density
$\rho_{s}$; namely, $\rho_{s}=\Delta N + K$ with the missing area
$\Delta N=8\int^{\omega_{c}}_{0^{+}}{}d\omega\left[\sigma^{n}_{1c}(\omega)
-\sigma^{s}_{1c}(\omega)\right]$ and $K=-4\pi e^{2}d
\left[\langle H_{c}\rangle^{s}-\langle H_{c}\rangle^{n}\right]$,
where the cutoff $(\omega_{c})$ is to be applied at the 
energy of the interband transitions and $\sigma^{n(s)}_{1c}(\omega)$
is the real part of the $c$-axis optical conductivity in normal(n) and 
superconducting(s) state, and $e$ and $d$ are the electron charge and
the interlayer spacing, respectivly.
The difference in $c$-axis kinetic energy between superconducting and normal
state $\left[\langle H_{c}\rangle^{s}-\langle H_{c}\rangle^{n}\right]$
gives the correction to the conventional sum rule for which 
the kinetic energy difference vanishes.

The effect of an in-plane magnetic
field on the $c$-axis kinetic energy difference can be expressed
in terms of the Green's function $G({\bf k},\omega)$
and the corresponding Gorkov function $F({\bf k},\omega)$.
We assume that the field penetrates freely into the sample
so that it is uniform between CuO$_{2}$ planes.\cite{kim2}
First, we consider the case of coherent $c$-axis coupling $t_{\perp}({\bf k})$.
The $c$-axis kinetic
energy is
\begin{eqnarray}
\langle H_{c}\rangle^{s}_{\bf q}=&&4T\sum_{\omega}\sum_{\bf k}
t_{\perp}({\bf k})^{2}
\left[G({\bf k}+{\bf q},{\tilde\omega})G({\bf k},{\tilde\omega})
-F({\bf k}+{\bf q},{\tilde\omega})F({\bf k},{\tilde\omega})\right]
\nonumber\\
\simeq&&-4TN(0)\sum_{\omega}\int{d\phi\over2\pi}t_{\perp}({\bf k})^{2}
\int^{\omega_{c}}_{-\omega_{c}} d\xi
{{{\tilde\omega}^{2}-\xi(\xi+\epsilon_{\bf q})+\Delta^{2}_{\bf k}}\over
{[{\tilde\omega}^{2}+\xi^{2}+\Delta^{2}_{\bf k}]
[{\tilde\omega}^{2}+(\xi+\epsilon_{\bf q})^{2}+\Delta^{2}_{\bf k}]}}\;,
\label{kinetic}
\end{eqnarray}
where
${\tilde\omega}=\omega+\gamma\mbox{sgn}\omega$ with fermionic
Mastubara frequency $\omega$, $\gamma$ is an effective impurity
scattering rate modeled as a constant for simplicity, 
and ${\bf q}=(ed/2){\hat {\bf z}}\times{\bf H}$
with ${\hat {\bf z}}$ a unit vector along the $c$-axis.
The Green's functions in Eq.~(\ref{kinetic}) with ${\bf k}+{\bf q}$ involve
the quasiparticle energy $E_{{\bf k}+{\bf q}}\simeq
\sqrt{\xi^{2}_{{\bf k}+{\bf q}}+\Delta^{2}_{\bf k}}$ with
$\xi_{{\bf k}+{\bf q}}\simeq\xi_{\bf k}+\epsilon_{\bf q}$.\cite{gap} 
Here,
$\xi_{\bf k}$ is energy spectrum measured with respect to the Fermi energy
and $\epsilon_{\bf q}=qv_{F}\cos(\phi-\theta_{q})$ with the Fermi
velocity $v_{F}$ and $\theta_{q}$ the direction of ${\bf q}$. Note that
$\theta_{q}=\theta+\pi/2$ can be interpreted as 
the direction of ${\bf H}\;(\theta)$ 
because of the symmetry in the problem. 

The difference $\langle H_{c}\rangle^{s}_{\bf q}-\langle H_{c}\rangle^{s}$
turns out to be, after some algebra,
\begin{equation}
\langle H_{c}\rangle^{s}_{\bf q}-\langle H_{c}\rangle^{s}\simeq
{N(0)\over3}\int{d\phi\over2\pi}t_{\perp}({\bf k})^{2}
\left({\omega_{c}\over T}\right)^{2}\tanh\left({\omega_{c}\over 2T}\right)
\left[\tanh\left({\omega_{c}\over 2T}\right)^{2}-1\right]
\left({\epsilon_{\bf q}\over\omega_{c}}
\right)^{2}\;.
\label{kdiff}
\end{equation}
The correction to Eq.~(\ref{kdiff}) is
${\cal O}\left[(v_{F}q/\omega_{c})^{2}(\Delta_{0}/\omega_{c})^{2}\right]$.
Note that $(v_{F}q/\omega_{c})^{2}$ is negligible and, furthermore,
the quantity in the square bracket vanishes
since $T\ll\omega_{c}$.
The same holds in the normal state {\it i.e.}
$\langle H_{c}\rangle^{n}_{\bf q}-\langle H_{c}\rangle^{n}\simeq0$.
Consequently, the kinetic energy difference between
superconducting and normal state does not change
with in-plane magnetic field.
For incoherent coupling, the same conclusion applies but for simpler reason. 
In this case
\begin{equation}
\langle H_{c}\rangle^{s}_{\bf q}
\simeq 4TN(0)\sum_{\omega}
\int{d\phi_{k}d\phi_{p}\over(2\pi)^{2}}|V(\phi_{k},\phi_{p})|^{2}
\int d\xi_{\bf k}d\xi_{\bf p}
\Phi\left(\xi_{\bf k}+\epsilon_{\bf q},\Delta_{\bf k}\right)
\Phi\left(\xi_{\bf p},\Delta_{\bf p}\right)\;,
\end{equation}
where $\Phi(\xi,\Delta)=\left(i{\tilde\omega}+\xi-\Delta\right)/
\left({\tilde\omega}^{2}+\xi^{2}+\Delta^{2}\right)$.
By changing $\xi_{\bf k}+\epsilon_{\bf q}\rightarrow\xi_{\bf k}$,
we see that $\langle H_{c}\rangle^{s}_{\bf q}=\langle H_{c}\rangle^{s}$,
and there is no change in kinetic energy.

The superfluid density $\rho_{s}$ is $\rho_{s}=\Delta N + K$.
In the presence of an in-plane field, the superfluid density 
$(\rho_{s,\bf q})$ becomes
$\Delta N_{\bf q}+K_{\bf q}$,
where $\rho_{s,{\bf q}}=\rho_{s}+\delta\rho_{s}$,
$\Delta N_{\bf q}=\Delta N+\delta\Delta N$, and
$K_{\bf q}=K+\delta K$. Since $\delta K=0$, we find
$\delta\rho_{s}=\delta\Delta N$.
The conductivity sum rule (or the normalized missing spectral weight)
is 
\begin{equation}
{\Delta N_{\bf q}\over\rho_{s,{\bf q}}}=
{\Delta N+\delta\Delta N\over
\rho_{s}+\delta\rho_{s}}\simeq
{\Delta N\over\rho_{s}}+{\delta\rho_{s}\over\rho_{s}}
\left[1-{\Delta N\over\rho_{s}}\right]\;.
\end{equation}
For pure coherent coupling, $\Delta N/\rho_{s}=1$, and for pure
incoherent coupling, $\delta\rho_{s}=0$; 
therefore, the sum rule is not changed by ${\bf H}$.
However, if both are present,
the sum rule is changed because $\delta\rho_{s}\ne0$ from the 
coherent contribution, and $\Delta N/\rho_{s}\ne1$
from the incoherent part.\cite{kim}

We next compute the change in superfluid density in the presence
of ${\bf H}$ only for
coherent coupling because ${\bf H}$ has no effect for
incoherent coupling.
For a constant $t_{\perp}$,
the superfluid density $\rho_{s,\bf q}$ is
$\rho_{s,\bf q}=t^{2}_{\perp}{\cal C}
T\sum_{\omega}\sum_{\bf k}F({\bf k}+{\bf q},{\tilde\omega})
F({\bf k},{\tilde\omega})$,
where ${\cal C}=16\pi e^{2}d$. Assuming a cylindrical Fermi surface,
we obtain, as $T\rightarrow0$, 
\begin{equation}
\rho_{s,\bf q}=t^{2}_{\perp}{\cal C}N(0)
\int{d\phi\over2\pi}{\Delta^{2}_{\bf k}\over \epsilon_{\bf q}
\sqrt{4\Delta^{2}_{\bf k}+\epsilon^{2}_{\bf q}}}
\ln\left[{\Psi^{(1)}_{+}\Psi^{(2)}_{-}\over
\Psi^{(1)}_{-}\Psi^{(2)}_{+}}\right]\;,
\end{equation}
where $\Psi^{(1)}_{\pm}=\sqrt{4\Delta^{2}_{\bf k}+\epsilon^{2}_{\bf q}}\pm
\epsilon_{\bf q}$ and
$\Psi^{(2)}_{\pm}=\sqrt{4\Delta^{2}_{\bf k}+\epsilon^{2}_{\bf q}}
\sqrt{\Delta^{2}_{\bf k}+\gamma^{2}}\pm\gamma \epsilon_{\bf q}$.
At finite $T$ with $\gamma=0$, we obtain
\begin{equation}
\rho_{s,\bf q}(T)=t^{2}_{\perp}{\cal C}\sum_{\bf k}
{\Delta^{2}_{\bf k}
\left[\chi(E_{\bf k})-\chi(E_{{\bf k}+{\bf q}})\right]
\over 2(E^{2}_{{\bf k}+{\bf q}}-E^{2}_{\bf k})}\;,
\label{rhoqT}
\end{equation}
where $\chi(E)=\tanh(E/2T)/E$.
For general coherent coupling,
we simply replace $t^{2}_{\perp}$ by
$t_{\perp}({\bf k})^{2}$ within the
sum over ${\bf k}$. 
As ${\bf H}\rightarrow0$, $\rho_{s,\bf q}(T)\rightarrow\rho_{s}(T)$.
Applying the nodal approximation\cite{lee} to the general case for
$T\ll\Delta_{0}$, we obtain, at ${\bf H}=0$, for a pure case
$\rho_{s}(T)=\rho_{s}(0)
-{\cal C}N(0)\left[\alpha_{1}t^{2}_{\perp}\left(T/\Delta_{0}\right)
+\alpha_{3}t_{\perp}t_{\phi}\left(T/\Delta_{0}\right)^{3}
+\alpha_{5}t^{2}_{\phi}\left(T/\Delta_{0}\right)^{5}\right]$,
where 
$\left[\alpha_{1},\alpha_{3},\alpha_{5}\right]=
\left[\ln(2),\;4.48\zeta(3),\;42.2\zeta(5)\right]$.
The $t^{2}_{\perp}$ contribution is linear in $T$ as expected and
the $t^{2}_{\phi}$ goes like $T^{5}$, a well known result.\cite{xiang1}
The $t_{\perp}t_{\phi}$ cross term gives $T^{3}$ just as does the impurity
mediated contribution.\cite{hirschfeld} 
In the above expression, $\rho_{s}(0)=
{\cal C}N(0)\left[t^{2}_{\perp}/2+t_{\perp}t_{\phi}/2
+3t^{2}_{\phi}/16\right]$.
From here on we will use the notation $\rho^{(i)}_{s,\bf q}$ with
$i=1$, $2$, and $3$ for $t^{2}_{\perp}$, $t_{\perp}t_{\phi}$,
and $t^{2}_{\phi}$ contribution, respectively.

Numerical results for $\rho^{(i)}_{s,\bf q}/\rho^{(i)}_{s}(0)$
as a function of $\epsilon_{q}/T_{c}$ 
with $\theta_{q}=0$ and $\gamma=0$ are given in Fig.~(1).
One should not confuse $\epsilon_{q}(\equiv qv_{F})$ with 
$\epsilon_{\bf q}=\epsilon_{q}\cos(\phi-\theta_{q})$.
The curve labeled $(1)$ refers to the $t^{2}_{\perp}$ contribution
and is nearly a straight line implying a linear dependence on $H$ 
(the magnitude
of ${\bf H}$).
For $t(\phi)^{2}=t^{2}_{\phi}\cos^{4}(2\phi)$, the effect of the field on 
$\rho^{(3)}_{s,\bf q}/\rho^{(3)}_{s}(0)$ is now much smaller
than for $t^{2}_{\perp}$
as the curve $(3)$ shows. This reduction results because
the nodal quasiparticles are eliminated
from the $c$-axis transport in this case by the factor
$\cos^{4}(2\phi)$. The numerical data for the $t^{2}_{\phi}$ case fit well
an $H^{2}$ law at small $H$. The last curve in Fig.~(1) which is labeled
by $(2)$ refers to the cross term $2t_{\perp}t_{\phi}\cos^{2}(2\phi)$.
It also fits an $H^{2}$.
Thus the constant $t_{\perp}$ part in $t_{\perp}({\bf k})$
is most effective in producing a change in $\rho_{s,\bf q}$
in the presence of ${\bf H}$.

It is instructive to derive an approximate, but analytic expression
at $T=0$, for the ${\bf H}$ dependence of $\rho_{s,\bf q}$
just discussed. We start from the temperature-independent part
of Eq.~(\ref{rhoqT}).
Applying a nodal approximation\cite{lee}
should be reasonable
for the difference between $\rho_{s,\bf q}$ and $\rho_{s}(0)$.
In this approximation, 
$\sum_{\bf k}\rightarrow\sum_{\mbox {node}}\int{\cal J} pdpd\vartheta$,
where ${\cal J}=\left[(2\pi)^{2}v_{F}v_{G}\right]^{-1}$
after an appropriate coordinate transformation.
The integration over $p$ is to be limited to the nodal region with cutoff
$p_{0}$ of order $\Delta_{0}$. 
The quasiparticle energy 
$E_{\bf k}=\sqrt{\xi^{2}_{\bf k}+\Delta^{2}_{\bf k}}$ then takes the form
$E_{\bf k}=\sqrt{p^{2}_{1}+p^{2}_{2}}=p$, where
$p_{1}=p\cos(\vartheta)$ and $p_{2}=p\sin(\vartheta)$.
In ${\cal J}$, $1/(\pi v_{F}v_{G})=N(0)/\Delta_{0}$ with $N(0)$ the 
density of states at the FS. Near a node
we can approximate 
$\epsilon_{\bf q}\simeq\epsilon_{q}\cos(\phi_{n}-\theta_{q})$,
where $\phi_{n}=\pi/4$, $3\pi/4$, $5\pi/4$, or $7\pi/4$. 
Thus for a give value of $\theta_{q}$, say $\theta_{q}=0$,
$\epsilon_{\bf q}\simeq\pm\epsilon_{q}/\sqrt{2}$.

For the $t^{2}_{\perp}$ case, 
choosing $\theta_{q}=0$, that is ${\bf H}$
along an anti-node, applying the nodal approximation we obtain
${\rho^{(1)}_{s,\bf q}/\rho^{(1)}_{s}(0)}
=1-(\pi/8\sqrt{2})
(\epsilon_{q}/\Delta_{0})$, 
where $\rho^{(1)}_{s}(0)=t^{2}_{\perp}{\cal C}N(0)/2$.
We choose $\Delta_{0}=2.14\;T_{c}$, then
$\rho^{(1)}_{s,\bf q}/\rho^{(1)}_{s}(0)=1-(\epsilon_{q}/T_{c})/7.71$.
Comparison with our numerical result given in Fig.~(1) shows that
this simple expression works well up to $\epsilon_{q}=0.5\;T_{c}$.
Similarly, for both $t(\phi)^{2}$ and $t_{\perp}t(\phi)$ term, we obtain
${\rho^{(i)}_{s,\bf q}/\rho^{(i)}_{s}(0)}
\simeq1-\left({\epsilon_{q}/\Delta_{0}}\right)^{2}/4$,
where $i=2$ and $3$.
Consequently, in the presence of
the in-plane magnetic field along the anti-node,
the $\epsilon_{q}$ dependence
of $\rho_{s,\bf q}$ is $|\epsilon_{q}|$ for the $t^{2}_{\perp}$ term, and
$\epsilon^{2}_{q}$ for both $t_{\perp}t(\phi)$ and $t(\phi)^{2}$ terms.
Even if ${\bf H}$ is in the nodal direction, the $\epsilon_{q}$ dependence
of $\rho_{s,\bf q}$ does not change; however, the coefficient of
$|\epsilon_{q}|$ (or
$\epsilon^{2}_{q}$) becomes smaller because the field has no effect on
the quasiparticle with $\phi_{n}\simeq\theta\pm\pi/2$.

While in Fig.~(1)
we took the field along the anti-nodal direction, in Fig.~(2)
we show additional results as a function of $\theta$ in the
range $0$ to $\pi$ for specific values of $\epsilon_{q}$.
The pattern obtained has four fold symmetry for a $d$-wave
gap. Note that the underlying symmetry of the problem
ensures that $\rho_{s,\bf q}$ is periodic in $\pi/2$
as is the quasiparticle conductivity.\cite{kim2}
The thin solid line is for a constant $t^{2}_{\perp}$ with
$\epsilon_{q}=0.1T_{c}$, $\gamma=0$, and $T=0$ while
the thick solid line is for $\epsilon_{q}=0.2T_{c}$ (higher magnetic field).
At all angle $\theta$, $\rho_{s,\bf q}$ is reduced over its zero field value.
There is a minimum along the anti-nodal direction and
a maximum along the nodal direction. 
In fact, for ${\bf H}$ along the node it can be shown by an application of the
nodal approximation that $\rho^{(1)}_{s,\bf q}/\rho^{(1)}_{s}(0)=
1-(\pi/16)\epsilon_{q}/\Delta_{0}$. 
Note that the ratio of anti-nodal to nodal reduction
in $\rho_{s,\bf q}$
is $\sqrt{2}$ which is verified in our complete
numerical calculations. This happens because the superfluid density
depends on $|\epsilon_{q}|$ for a constant hopping amplitude.

Temperature and impurities will have an effect on the four fold
symmetry pattern shown in Fig.~(2). The first short dashed curve is for
$\rho^{(1)}_{s,\bf q}(T)$ with $\gamma=0$ but now $T=0.001\;T_{c}$ while
the second is for $T=0.01\;T_{c}$. Both are to be compared with 
the $T=0$ result
(thin solid curve). We see that in the first case $T$ simply
rounds off a little the maxima in the nodal directions while
for a higher $T=0.01\;T_{c}$, which is still much less than
the magnetic energy $\epsilon_{q}=0.1\;T_{c}$, the temperature
effects are more significant and much of the anisotropy is washed out.
To see the anisotropy, which comes only from $t^{2}_{\perp}$,
one clearly needs to go to low temperature compared to the magnetic energy.
Impurities also affect the
anisotropy of $\rho_{s,\bf q}$ and results for $\gamma=0.002\;T_{c}$
are shown as the
long dashed curve in Fig.~(2). These results are to be compared with
the thick solid curve which applies for
the same value of $\epsilon_{q}=0.2\;T_{c}$. Impurities reduces the $c$-axis
superfluid density as we expect, and smears out the anisotropy.
The final dot-dashed line in Fig.~(2) is for the term $t(\phi)^{2}$
with $\epsilon_{q}=0.1\;T_{c}$ and $\gamma=0$. The superfluid reduction is 
small in comparison to the other curves and shows no anisotropy.

More information on the effect of temperature, impurities
and the magnetic field on the superfluid density for the $t^{2}_{\perp}$
case is given in Fig.~(3). All curves are for the field along the
anti-node (except the solid curve in the lower inset, which
is for the field along the node).
The solid curve is for comparison and gives the $T$ dependence
of $\rho^{(1)}_{s}(T)/\rho^{(1)}_{s}(0)$ with ${\bf H}=0$. The
other curves are for different increasing values of 
$\epsilon_{q}/T_{c}=0.1$ (dashed), $0.2$ (dot-dashed), 
and $0.3$ (long dashed curve). At low $T$, the power law in $T$ is
modified going to a $T^{3}$ while at high $T$ it returns to $T$.
It is seen that $T$ 
always decreases the superfluid density as does $\epsilon_{q}$. 
In all our finite $T$ numerical results,
we have approximated the $T$ dependence of the gap amplitude
$\Delta(T)$ by $\Delta_{0}\tanh\left[1.74\sqrt{(T/T_{c})-1}\;\right]$.
In the 
top inset we show similar results but as functions of magnetic 
energy $\epsilon_{q}/T_{c}$ for different values of $\gamma/T_{c}$ at $T=0$.
The solid curve is for $\gamma/T_{c}=0$, with which we compare
other curves for $\gamma/T_{c}=0.01$ (dashed), $0.02$ (dot-dashed),
and $0.03$ (long dashed curve). Impurities, like $T$, have their greatest
effect at $\epsilon_{q}=0$.

We can obtain analytic results for the $T$ dependence of 
$\rho_{s,\bf q}(T)=\rho_{s,\bf q}+\delta\rho_{s,\bf q}(T)$. 
We start with the constant $t^{2}_{\perp}$ case. Applying the
nodal approximation already described, when the field is along 
an anti-node we obtain
${\delta\rho^{(1)}_{s,\bf q}(T)/\rho^{(1)}_{s}(0)}=
-12\zeta(3)\left({T/\Delta_{0}}\right)
\left({T/\epsilon_{q}}\right)^{2}$.
For the $t(\phi)^{2}$ term, we obtain
${\delta\rho^{(3)}_{s,\bf q}(T)/\rho^{(3)}_{s}(0)}\simeq
-9450\zeta(7)\left({T/\epsilon_{q}}\right)^{2}
\left({T/\Delta_{0}}\right)^{5}$,
which goes like $T^{7}$. Similarly, we have, for $t_{\perp}t(\phi)$,
${\delta\rho^{(2)}_{s,\bf q}(T)/\rho^{(2)}_{s}(0)}\simeq
-270\zeta(5)\left({T/\epsilon_{q}}\right)^{2}
\left({T/\Delta_{0}}\right)^{3}$, which gives $T^{5}$.
For incoherent coupling, $\rho_{s,\bf q}(T)=\rho_{s}(T)$ and
its dependence remains $T^{3}$ as shown in Ref.\cite{hirschfeld}
For coherent coupling, ${\bf H}$ along an anti-node
changes the $T$ dependence of the superfluid density
as follows: $\rho_{s,\bf q}(T)\sim T^{2}\rho_{s}(T)$.
If $\theta\simeq\phi_{n}$, say $\theta\simeq\pi/4$, then
$\epsilon_{\bf q}\simeq0$ for $\phi_{n}=3\pi/4,\;7\pi/4$. This means that
the in-plane field has no effect on quasiparticles with
$\phi_{n}=3\pi/4,\;7\pi/4$.
Consequently, $\rho_{s,\bf q}(T)\sim \rho_{s}(T)$ for ${\bf H}$
along a node. This is shown in the lower inset of Fig.~(3). The solid
curve is for the nodal direction while the dashed curve
is along the anti-node. The anisotropy
disappears for $T>0.1\;T_{c}$.

It is clear from our analysis that a contribution to $\rho_{s,\bf q}$
from a small $t^{2}_{\perp}$ term will lead to 
anisotropy in $\rho_{s,\bf q}$ as a function of the orientation of 
${\bf H}$, and be linear in $H$ for small field.
In the pure case, its dependence is linear in $T$
going over to $T^{3}$ at finite field along an anti-node with
$T <\epsilon_{q}$(magnetic energy). These characteristic variation are quite 
distinct from those predicted for a $t(\phi)^2$ term or for the cross
term $t_{\perp}t(\phi)$. In both these cases there is no anisotropy
at $T=0$ and the changes with ${\bf H}$ are quadratic
$(\epsilon^{2}_{q})$. In the pure case with zero field,
the $T$ variations are $T^{5}$ and $T^{3}$, respectively, and go over to
$T^{7}$ and $T^{5}$ at finite field for $T<\epsilon_{q}$. These
characteristic laws should help constraint further the
relative magnitude of $t_{\perp}$, $t_{\phi}$ and also of incoherent
coupling. 
In this last instance, 
there is no effect of ${\bf H}$ and the $T^{3}$ remains at
finite field and low $T$.

In conclusion, we have considered the effect of an in-plane magnetic field
${\bf H}$ on the $c$-axis sum rule and superfluid density$(\rho_{s,\bf q})$.
We have assumed that the in-plane dynamics
can be described within a Fermi liquid theory. If the $c$-axis
coupling is coherent,
the sum rule remains conventional and unaffected by the
application of ${\bf H}$ but $\rho_{s,\bf q}$ itself is
reduced with increasing ${\bf H}$. The reduction with
${\bf H}$ depends on the exact functional form of the $c$-axis
coherent coupling and is different for a constant matrix element
$t_{\perp}$ and $t(\phi)=t_{\phi}\cos^{2}(2\phi)$. 
In the first instance, 
to leading order, it varies like $H$ (magnitude of ${\bf H}$) while
in the second like $H^{2}$. 
If, however, the $c$-axis transport is incoherent, then
both the sum rule and $\rho_{s,\bf q}$
remains unchanged. 
When both coherent and incoherent
interlayer transfer contribute, both the sum rule and $\rho_{s,\bf q}$
will change with ${\bf H}$. The change in $\rho_{s,\bf q}$
comes only from the coherent part.

For the case of coherent $c$-axis coupling with a constant $t_{\perp}$,
the change in $\rho_{s,\bf q}$ depends on the direction of ${\bf H}$
with respect to the nodes. For ${\bf H}$ along an anti-nodal direction
$\rho_{s,\bf q}$ is minimum while it has a sharp maximum when
${\bf H}$ is along a nodal direction. This anisotropy in
$\rho_{s,\bf q}$ is reduced when impurities are present or when
the temperature is increased. For the case of $t(\phi)$, the contribution 
to the transport coming from the nodes 
is eliminated  and the anisotropy disappears. The incoherent coupling
contribution also gives an isotropic response because in this case
momemtum is not conserved and, correspondingly, the momentum shift
caused by ${\bf H}$ is immaterial.

\begin{figure}

\caption
{$\rho^{(i)}_{s,{\bf q}}/\rho^{(i)}_{s}(0)$ as functions of
$\epsilon_{q}/T_{c}$ for $T=0$, $\theta=0$ and $\gamma=0$.
Note that the superscripts $(1)$, $(2)$ and $(3)$ represent
$t^{2}_{\perp}$, $t_{\perp}t_{\phi}$ and $t^{2}_{\phi}$, respectively.
}
\caption
{$\rho^{(i)}_{s,{\bf q}}(T)/\rho^{(i)}_{s}(0)$ as functions of
$\theta$. The dot-dashed line is for $t^{2}_{\phi}$
with $\epsilon_{q}=0.1T_{c}$, $\gamma=0$, and $T=0$.
The thin solid curve is for $t^{2}_{\perp}$ with
$\epsilon_{q}=0.1T_{c}$, $\gamma=0$, and $T=0$.
The two dashed curves just below this curve are for $t^{2}_{\perp}$ with
$\gamma=0$, and $T=0.001T_{c}$ and $T=0.01T_{c}$, respectively.
The thick solid curve is for the case of $t^{2}_{\perp}$ with
$\epsilon_{q}=0.2T_{c}$, $\gamma=0$, and $T=0$.
The long-dashed curve is for $\epsilon_{q}=0.2T_{c}$, $\gamma=0.002T_{c}$,
and $T=0$.
 }
\caption
{$\rho^{(1)}_{s,{\bf q}}/\rho^{(1)}_{s}(0)$ as functions of
$T/T_{c}$ for different values of $\epsilon_{q}/T_{c}$
with $\theta=0$ and $\gamma=0$ (pure case).
The solid curve is for $\epsilon_{q}/T_{c}=0$ (zero field).
The dashed, dot-dashed, and long-dashed curve for
$\epsilon_{q}/T_{c}=0.1$, $0.2$, and $0.3$, respectively.
In the upper inset,
$\rho^{(1)}_{s,{\bf q}}/\rho^{(1)}_{s}(0)$ is given as functions of
$\epsilon_{q}/T_{c}$ for different values of $\gamma/T_{c}$
with $\theta=0$ and $T=0$.
The solid, dashed, dot-dashed, and long-dashed curve for
$\gamma/T_{c}=0$, $0.01$, $0.02$, and $0.03$, respectively.
In the lower inset,
$\rho^{(1)}_{s,{\bf q}}/\rho^{(1)}_{s}(0)$ as functions of
$T/T_{c}$ for $\epsilon_{q}=0.3\;T_{c}$. The solid curve is for
$\theta\simeq\pi/4$ and the long-dashed curve for $\theta=0$.
}
\end{figure}


\begin{references}
\bibitem{gap} The gap seen by quasiparticles with ${\bf k}+{\bf q}$
is not $\Delta_{\bf k}$ but $\Delta_{{\bf k}+{\bf q}}$; however,
this difference can be neglected in our calculation.
See Ref.\cite{kim2}
\end{references}
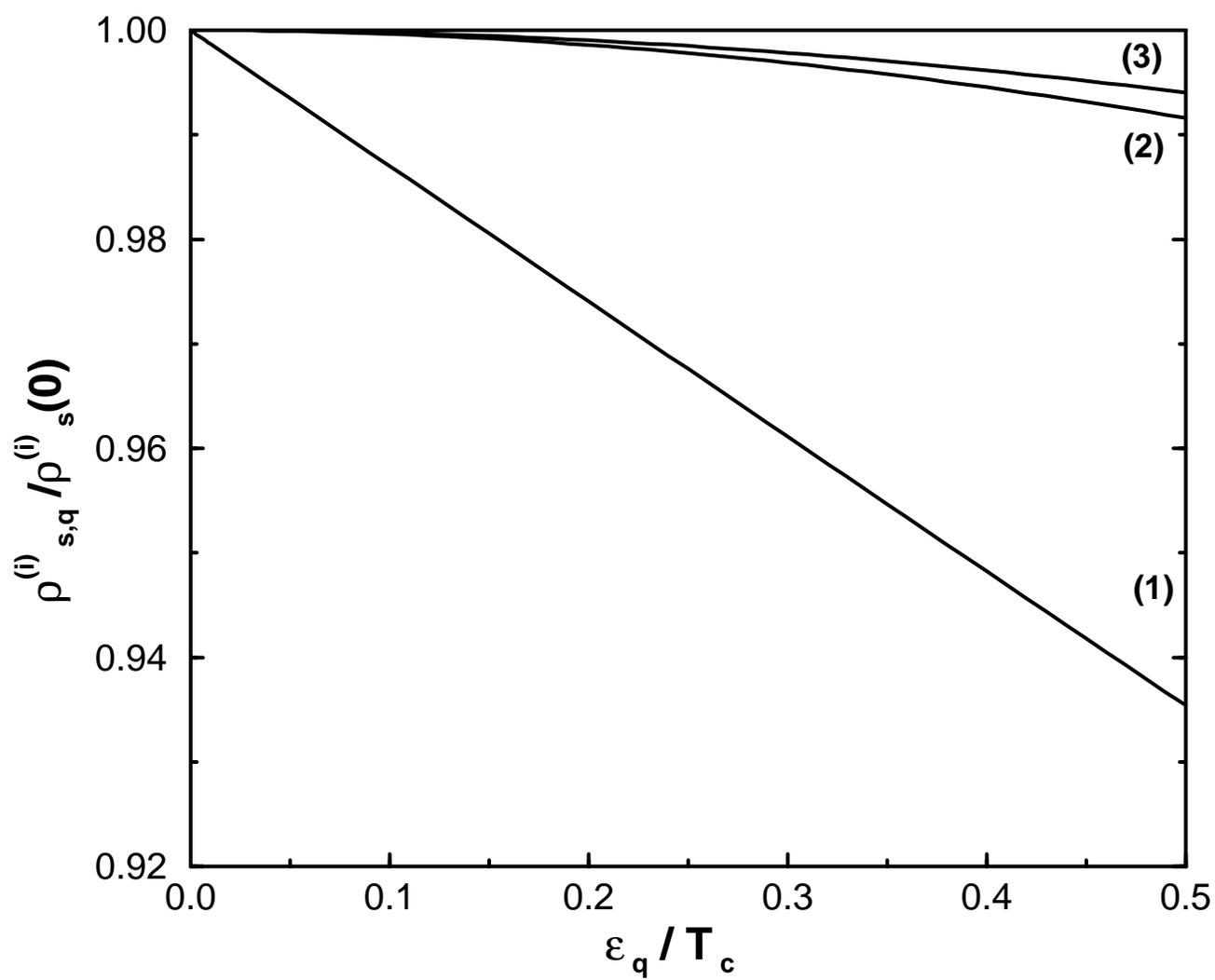
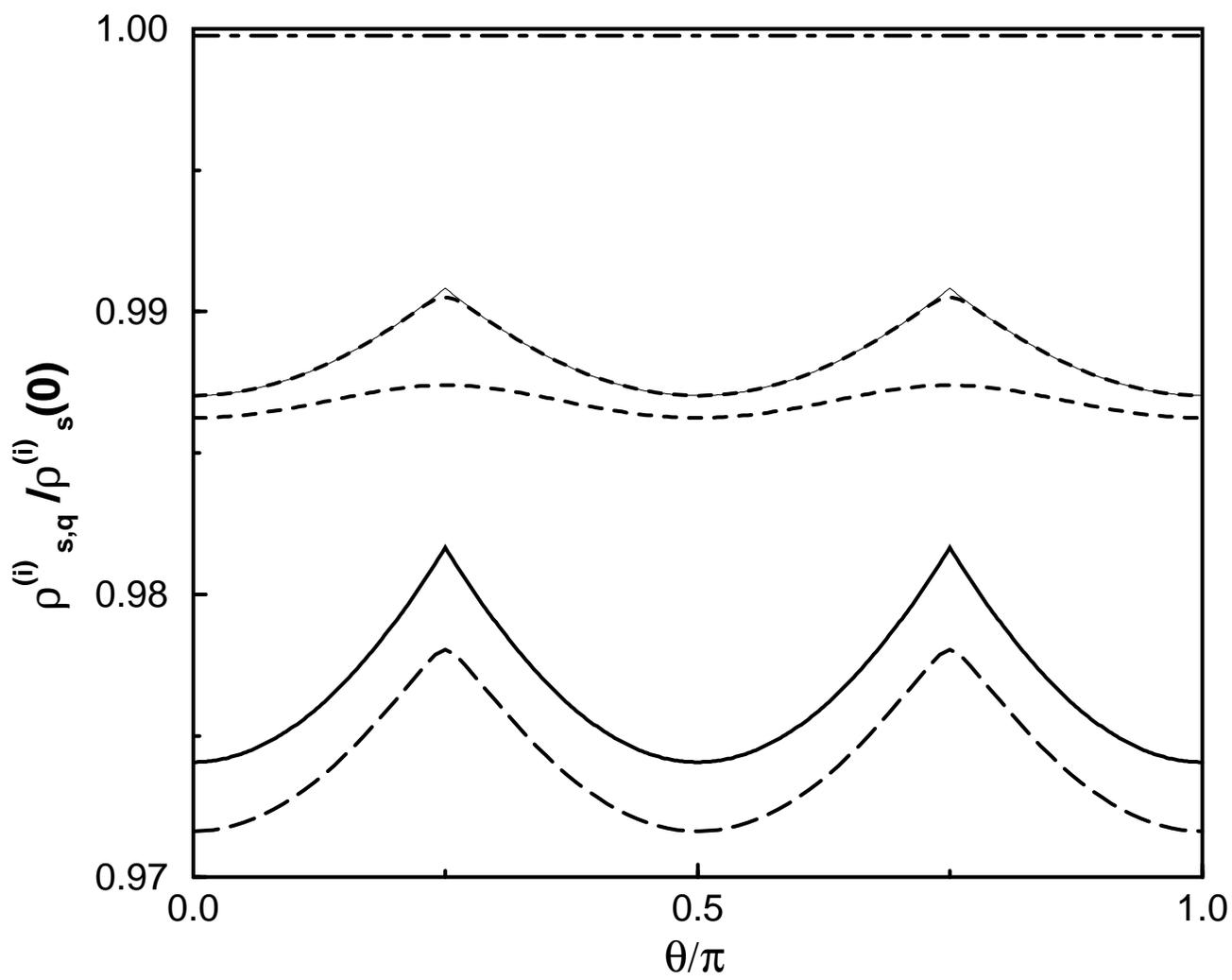
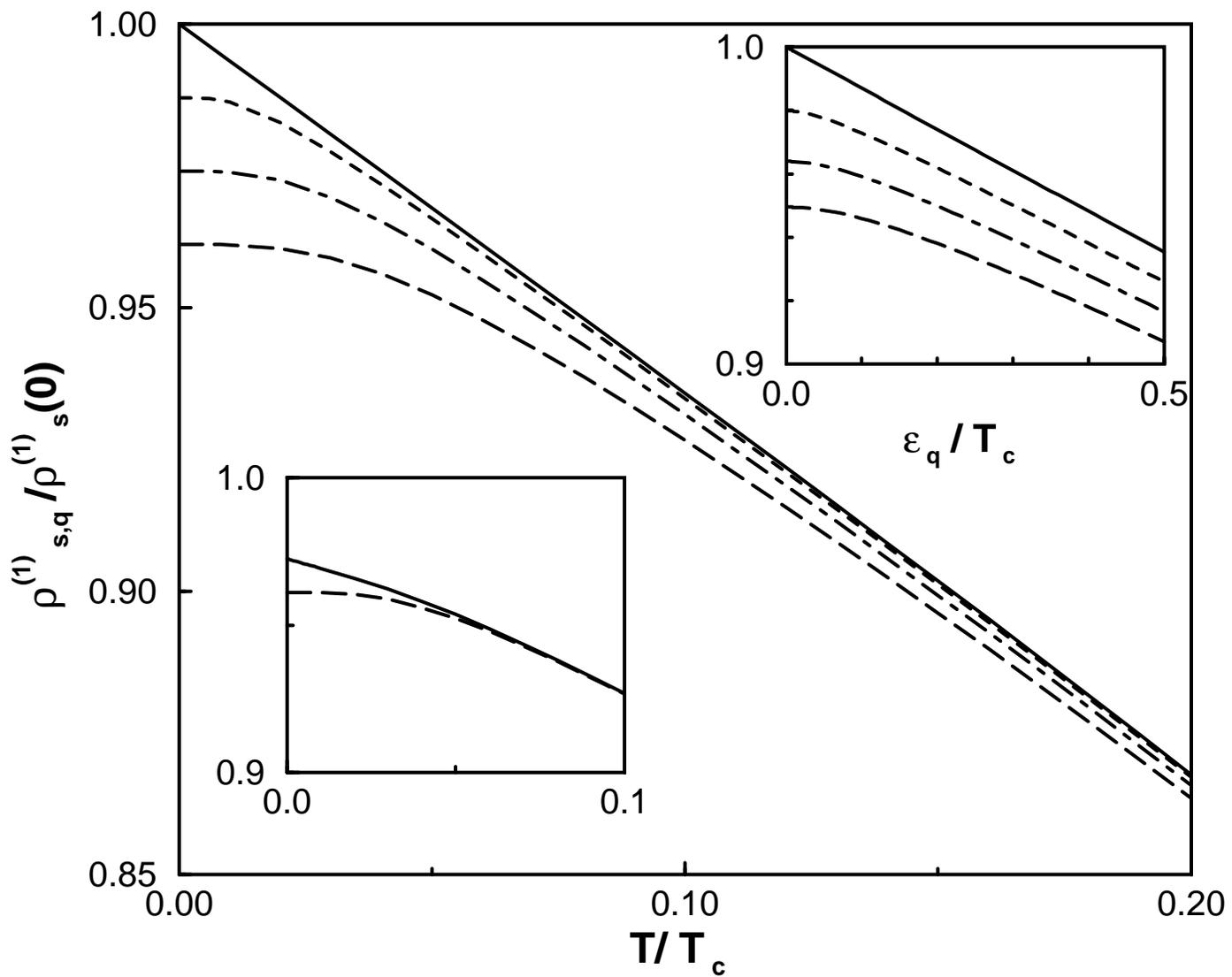
\end{document}